\documentclass[12pt]{spieman}  

\usepackage{amsmath,amsfonts,amssymb}
\usepackage{graphicx}
\usepackage{setspace}
\usepackage{tocloft}
\usepackage{lineno}
\usepackage{mathrsfs}
\usepackage{aas_macros}
\usepackage{float}
\usepackage{caption}
\usepackage{subcaption}
\usepackage{upgreek}
\usepackage[T1]{fontenc}

\usepackage[colorlinks=false]{hyperref}

\usepackage{fontawesome5}

\title{The \textsc{Toliman} mission: A low-cost space telescope for high precision narrow-angle astrometry}

\author[a, *]{Peter Tuthill}
\author[a]{Christopher Betters}
\author[a]{Max Charles}
\author[a]{Fred Crous}
\author[b]{Donald G. Dansereau}
\author[a,c]{Conaire Deagan}
\author[a,d]{Louis Desdoigts}
\author[a]{Mark George}
\author[a]{Thomas Holland}
\author[a,b]{Connor J. Langford}
\author[a]{Milo Langker}
\author[e]{Kieran Larkin}
\author[a]{Clarissa Luk}
\author[a]{Jack Nelson}
\author[f]{Benjamin Pope}
\author[a]{Grace Piroscia}
\author[a]{Angus Rutherford}
\author[a]{David Sweeney}
\author[a]{Adam Taras}
\author[a]{Karel Valenta}
\author[g]{Tim White}
\author[a, h]{Alison Wong}
\author[i]{Eduardo Bendek}
\author[d]{David Doelman}
\author[j]{Kyran Grattan}
\author[k]{Olivier Guyon}
\author[j]{Peter Klupar}
\author[c]{Benjamin T. Montet}
\author[l]{Jeffrey Smith}
\author[l]{Douglas Caldwell}
\author[i]{Frans Snik}
\author[j]{Simon P. Worden}

\affil[a]{Sydney Institute for Astronomy, School of Physics, The University of Sydney, NSW 2006, Australia}
\affil[b]{Australian Centre for Robotics, School of Aerospace, Mechanical and Mechatronic Engineering, The University of Sydney, NSW 2006, Australia}
\affil[c]{School of Physics, University of New South Wales, NSW  2052, Australia}
\affil[d]{Universiteit Leiden Sterrewacht, Einsteinweg 55, 2333 CC Leiden, Zuid-Holland, Netherlands}
\affil[e]{Nontrivialzeros Research, 22 Mitchell Street, Putney, NSW 2112, Australia}
\affil[f]{School of Mathematical and Physical Sciences, Macquarie University, Sydney, NSW 2109, Australia}
\affil[g]{Sydney Informatics Hub, Core Research Facilities, University of Sydney, NSW 2006, Australia}
\affil[h]{Discipline of Business Analytics, The School of Business Analytics and Marketing, The University of Sydney, NSW 2006, Australia}
\affil[i]{NASA Ames Research Center,  Moffett Field, CA 94035, USA}
\affil[j]{Breakthrough Prize Foundation, 9 Rue du Laboratoire, L-1911, Luxembourg}
\affil[k]{Subaru Telescope, National Observatory of Japan, 650 N. A'ohoku Place, Hilo, HI 96720 USA}
\affil[l]{SETI Institute, 339 Bernardo Ave, Suite 200, Mountain View, CA 94043, USA}

\usepackage{acronym}		
\usepackage{glossaries}		

\usepackage{booktabs}

\newacronym{USyd}{USyd}{the University of Sydney}

\newacronym{CAD}{CAD}{computer aided design}

\cftpagenumbersoff{figure}
\cftpagenumbersoff{table}

\begin{document}
\maketitle

\begin{abstract}

The \textsc{Toliman} project is engaged with the construction, launch and operation of a low-cost space telescope of unorthodox optical design. Its primary science goal targets an exhaustive search for temperate-orbit rocky planets around either star in the $\upalpha$~Centauri~AB binary within our nearest-neighbor star system. Despite their favorable proximity and brightness, the detection of terrestrial exoplanets around such nearby Sun-like stars remains problematic for contemporary instrumental approaches. By performing narrow-angle astrometric monitoring of binary stars at extreme precision, any exoplanets will betray their presence by way of gravitationally-induced perturbations on the binary orbit. 
Recovery of this signal is challenging for it amounts to only a few microarcseconds of angular deflection (at best), and so is normally thought to require a large (meter-class) instrument. 
By implementing an innovative optical and signal encoding architecture, the \textsc{Toliman} space telescope aims to recover such signals with a telescope aperture of only 12.5\,cm. 
This paper gives an overview of key features of the mission; in particular the concepts underlying the optics to enable image registration at the extreme levels of precision required. 
An outline is also provided, sketching further mission components and systems incorporated into the 16U CubeSat spacecraft bus in which the science payload is housed -- all of which are now under construction.
\end{abstract}

\keywords{Astrometry, Binary stars, Coded apertures, Diffractive optical elements, Exoplanetary science, Exoplanets, Image processing, Precision calibration, Precision optics, Space telescopes}

{\noindent \footnotesize\textbf{*} Correspondence to: Peter Tuthill,  \linkable{peter.tuthill@sydney.edu.au} }

\section{Introduction}

With the harnessing of ever more powerful technologies, the rapid growth in the rate of exoplanetary discovery (catalogs of confirmed objects are now climbing towards 10,000 entries \cite{Christiansen2025}) delivered by contemporary astronomy masks a pervasive observational blind spot: we are poorly equipped to study what is arguably the most compelling population of all -- nearby terrestrial exoplanets. 
For example, if we wish to know whether a Sun-like star hosts a rocky planet in a temperate orbit (a configuration yielding a {\it true Earth analog}), existing telescopes are not able to provide incisive answers. 
Photometric signals from planetary transits -- astronomy's most prolific discovery technique -- rely on an alignment between the plane of orbit and our line of sight that is highly unlikely ($<1$\,\%) to occur for Earth analog systems so that detections require monitoring of large populations which, perforce, lie at substantial  galactic distances\cite{Perryman2018}.
The second most prolific discovery technique, spectroscopic detection of radial velocity signals, although not requiring fortuitous alignment, has proved difficult to push to the precision to the levels required for Earth-mass objects orbiting at temperate-zone AU scales. 
Limiting noise floors arising from intrinsic stellar processes mean that straightforward improvements to instrumentation alone do not provide an immediate way forward \cite{Plavchan2015,Hara2023}.

As a consequence, astronomers' planetary census of Sun-like (FGK) stars is very incomplete for objects with masses below the realm of gas giants and few are known in our immediate stellar neighborhood (within 10\,pc). 
So while missions such as \textit{Kepler}\cite{Borucki2010} and the Transiting Exoplanet Survey Satellite (TESS)\cite{Ricker2015} do yield robust data on the statistical incidence of planets in various settings, including the near-solar region, we have scant specific details or confirmed detections of nearby Earth analog worlds.
Despite this, the discovery of such objects appears as a critical goal at the top rung of science priority for astronomy planning and roadmapping exercises worldwide. 
As the era of detection gives way to the science of exoplanetary characterisation, the critical questions concern the diversity of planetary chemistry, atmospheric composition and physics\cite{JontofHutter2019,Owen2019,Wordsworth2022}, surface abundances and potential for an exoplanetary biosphere \cite{Krissansen-Totton2018,Fujii2018}. 
In this task, our plentiful sample of exoplanets is confronted with a problem: most are at great distance ($\gtrapprox$ 100\,pc) making it problematic to perform follow-up observations such as adaptive optics imaging to isolate the faint planetary light for study away from the glare of the host star. 
On the other hand, planets (if they could be found) around nearby stars would be at ideal scales for this purpose, potentially yielding spectra in which the fingerprints for molecular chemistry or even biomarkers might be found. 

\begin{figure}[hbt!]
\centering
\includegraphics[width=15cm]{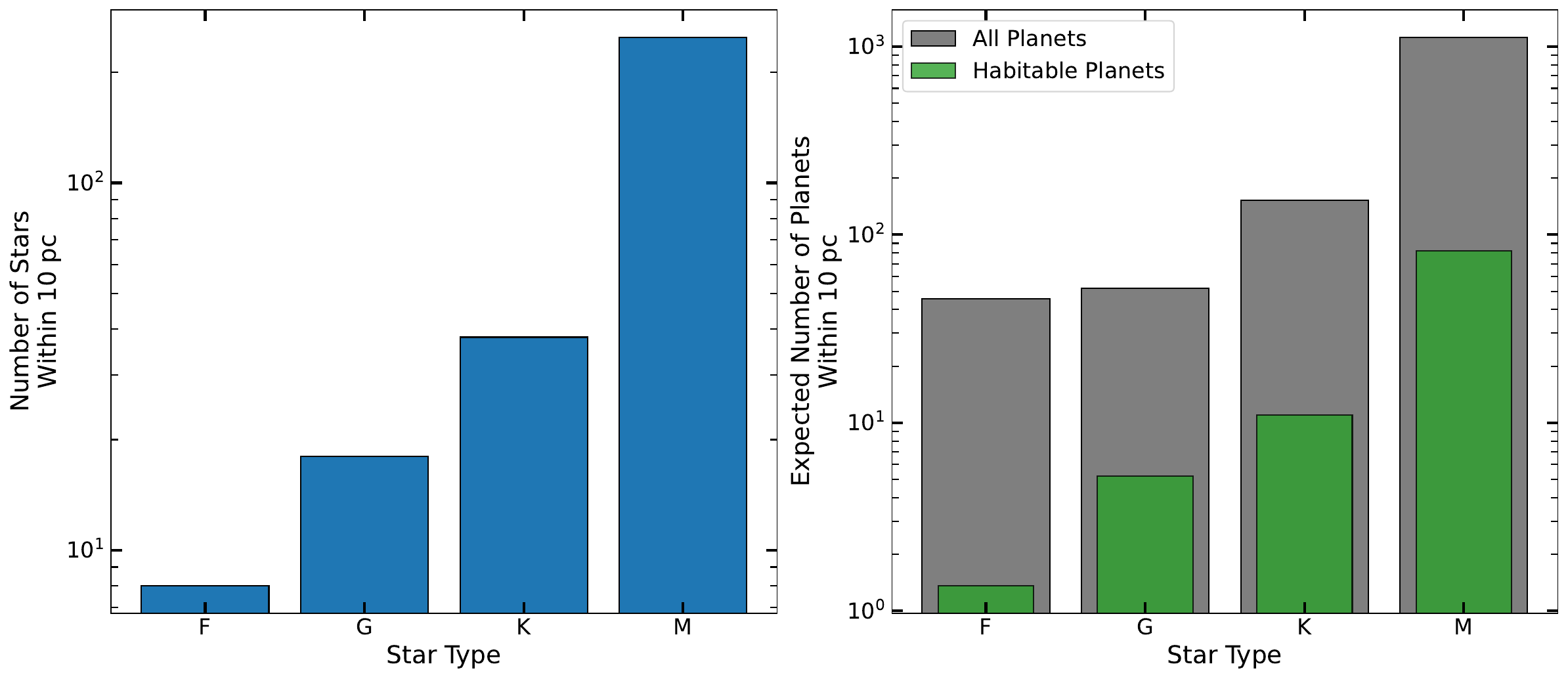}
\caption{Stellar demographics and expected planetary populations within 10 pc. \textit{Left:} Number of stars by spectral type\cite{Reyl2021}. \textit{Right:} Expected number of planets (gray) and habitable planets (green) by host star spectral type, calculated using median occurrence rates from Figure~\ref{fig:planet_num_estimation}.}
\label{fig:starcounts}
\end{figure}

\begin{figure}[hbt!]
\centering
\includegraphics[width=15cm]{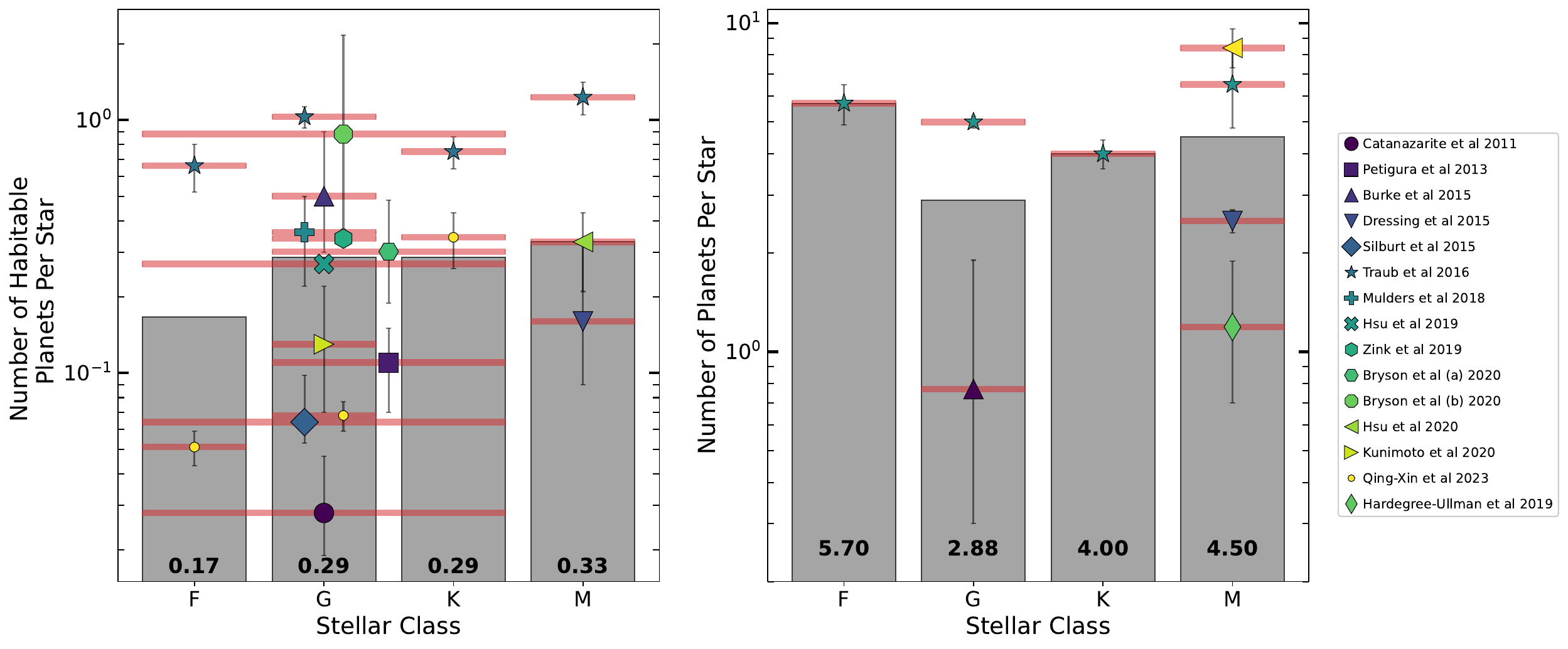}
\caption{Visual summary of planet occurrence rates by stellar class from selected literature. Gray bars show median literature values for each stellar class. Individual markers represent previous studies, with error bars indicating reported uncertainties\textsuperscript{$\dagger$} in occurrence estimates. Red shaded regions denote the range of stellar classes to which each estimate applies. Marker colors correspond to publication year, and some markers are horizontally offset for visual clarity. \textit{Left:} Average number of habitable\textsuperscript{$\ddagger$} planets per star. \textit{Right:} Average total number\textsuperscript{$\nmid$} of planets per star. $\dagger$ Refer to individual studies for specific uncertainty measures. $\ddagger$ Refer to individual studies for definitions of the habitable zone. $\nmid$ See individual studies for sensitivity limits and corrections.} 

\label{fig:planet_num_estimation}
\end{figure}

Being nearby and therefore relatively bright, the population of stars within a sphere with a radius of 10\,pc from the Sun is well known\cite{Kirkpatrick2024} and furthermore a 1\,AU star-planet orbital separation placed at this distance subtends an angle of 100\,milliarcseconds: well suited to recovery of high contrast companions with modern adaptive optics and coronagraph technologies implemented at present and coming-generation ground-based large aperture telescopes. 
The histogram of \autoref{fig:starcounts} breaks the 10\,pc stellar sample down by spectral type\cite{Reyl2021}, illustrating that our stellar neighborhood is predominantly populated by M-type stars and showing the rapidly falling mass function to higher mass Sun-like stars.
Published estimates of habitable planet occurrence rates span approximately two orders of magnitude, partially due to differing approaches to completeness corrections, stellar sample selection, and habitability criteria. Moreover, occurrence rate estimates at longer orbital periods are acutely sensitive to a small number of confirmed detections\cite{litrev_Bryson2020b}. Even a single confirmed detection of a habitable-zone planet around a star like $\alpha$ Centauri would provide a qualitatively different constraint on the statistical extrapolation, and thus help improve occurrence rate estimates. Due to the aforementioned factors, we adopted the median as a conservative central estimate. Combining this stellar distribution with median planet occurrence rates from a representative sample of the literature\cite{litrev_Bryson2020a, litrev_Bryson2020b, litrev_Burke2015, litrev_Catanzarite2011, litrev_Dressing2015, litrev_HardegreeUllman2019, litrev_Hsu2019, litrev_Hsu2020, litrev_Kunimoto2020, litrev_Mulders2018, litrev_Petigura2013, litrev_Qingxin2023, litrev_Silburt2015, litrev_sum_Biazzo2022, litrev_sum_Winn2015, litrev_traub2016, litrev_Zink2019}—approximately 0.25 habitable planets per FGK star—we estimate the expected planetary populations within 10 pc (\autoref{fig:starcounts}, right). This analysis yields statistical predictions for both the total exoplanet population and the subset of temperate rocky planets around F through M-type stars. We note that published estimates are in better agreement for short-period planets where detection samples are large and completeness corrections are well constrained. The scatter in \autoref{fig:planet_num_estimation} is driven by the long period habitable-zone regime. As the innermost edge of the habitable zone for $\alpha$ Centauri B corresponds to $\sim140$ days, the parameter space for the TOLIMAN targets falls squarely within this poorly constrained regime.
Considering a $<$10\,pc volume limited sample of stars of spectral types F, G and K (66 stars in total) to be ``Sun-like'', then there should exist roughly $\sim$15 associated exoplanets that would then qualify as ``true Earth analogs''.
Uncovering members of this unknown population is a key goal motivating the development and maturation of astrometric technologies, of which our \textsc{Toliman} mission is intended to be a pathfinder.

\section{Indirect detection by gravitational perturbations}

\subsection{Detection of companions by radial velocity monitoring}

For a two-body system, the gravitational reflex motion induced on a star by an orbiting planet yields signals that enable two of the major ``indirect'' methods for exoplanet detection: monitoring for variations in radial velocity (RV) and astrometric position.
Considering firstly the former of these, the component of this motion along the observer's line of sight induces cyclic shifts in apparent velocity with a semi-amplitude $K$ given by the equation\cite{Perryman2018}
\begin{equation}
    K = \left(\frac{2\pi G}{P}\right)^{1/3} \frac{M_p \sin(i)}{(M_*+M_p)^{2/3} (1-e^2)^{1/2}}
    \label{eq-rv}
\end{equation}
where $M_p$ and $M_*$ are the masses of the planet and star respectively, $i$, $P$ and $e$ are the inclination, period and eccentricity of the planet, and $G$ is the gravitational constant. 
It can be seen that the RV signal strength varies inversely with period and stellar mass ($P^{1/3}$ and $M_*^{2/3}$ from Equation~\ref{eq-rv}). 
Signals are therefore strongest for planets in short-period orbits around low-mass stars: a configuration ideally suited to probe close-in temperate zone orbits around low-luminosity M~dwarf stars.

On the other hand, these same systematics make RV less sensitive to the significantly longer $\sim$year orbits \cite{Perryman2018} associated with the habitable zone around more luminous solar-analog stars. 
The strength of RV signals induced by an Earth-mass planet placed in a temperate orbit, calculated over a range of stellar types, is shown in Figure~\ref{fig_signalstrength}. 
This illustrates that the strength of such signals falls below a few tens of cm/sec for stars hotter than M-class, so that prospects for clear detections of a true Earth analog by way of radial velocity monitoring confront a forbidding observational challenge. 
Current instrumental noise floors factoring extremely stable spectrograph design are not far from such levels of precision\cite{Espresso_2021}, however the problem is greatly compounded by the presence of noise processes intrinsic to the stellar atmospheres themselves so that improvements to the technology alone will not necessarily advance detection thresholds to the required goals.

\subsection{Detection of companions by astrometric monitoring}

\label{sec:astro_mon}

In common with the previously discussed RV method, astrometry is ``indirect'' in that it does not witness the exoplanet itself. 
Rather, it relies on precise positional measurements of the host star over a period of time in which small variations reveal the existence and properties of an unseen companion. 
The stellar motion around the barycentre of the two-body (star-planet) system will be witnessed as a small elliptical locus to a distant observer. 
That component of the gravitational reflex motion in the plane of the sky induces a positional angular perturbation that enables the class of astrometric detection strategies.
The angular deflection on the sky $\phi$ (in arcseconds) for a system at distance $d$ (in pc) with orbit semi-major axis $a$ (in AU) is given by \cite{Perryman2018}

\begin{equation}
    \phi = \left(\frac{M_p}{M_*}\right)\left(\frac{a}{d}\right).
    \label{eq-astrom}
\end{equation}

\begin{figure*}
\centering
\includegraphics[width=0.8\linewidth]{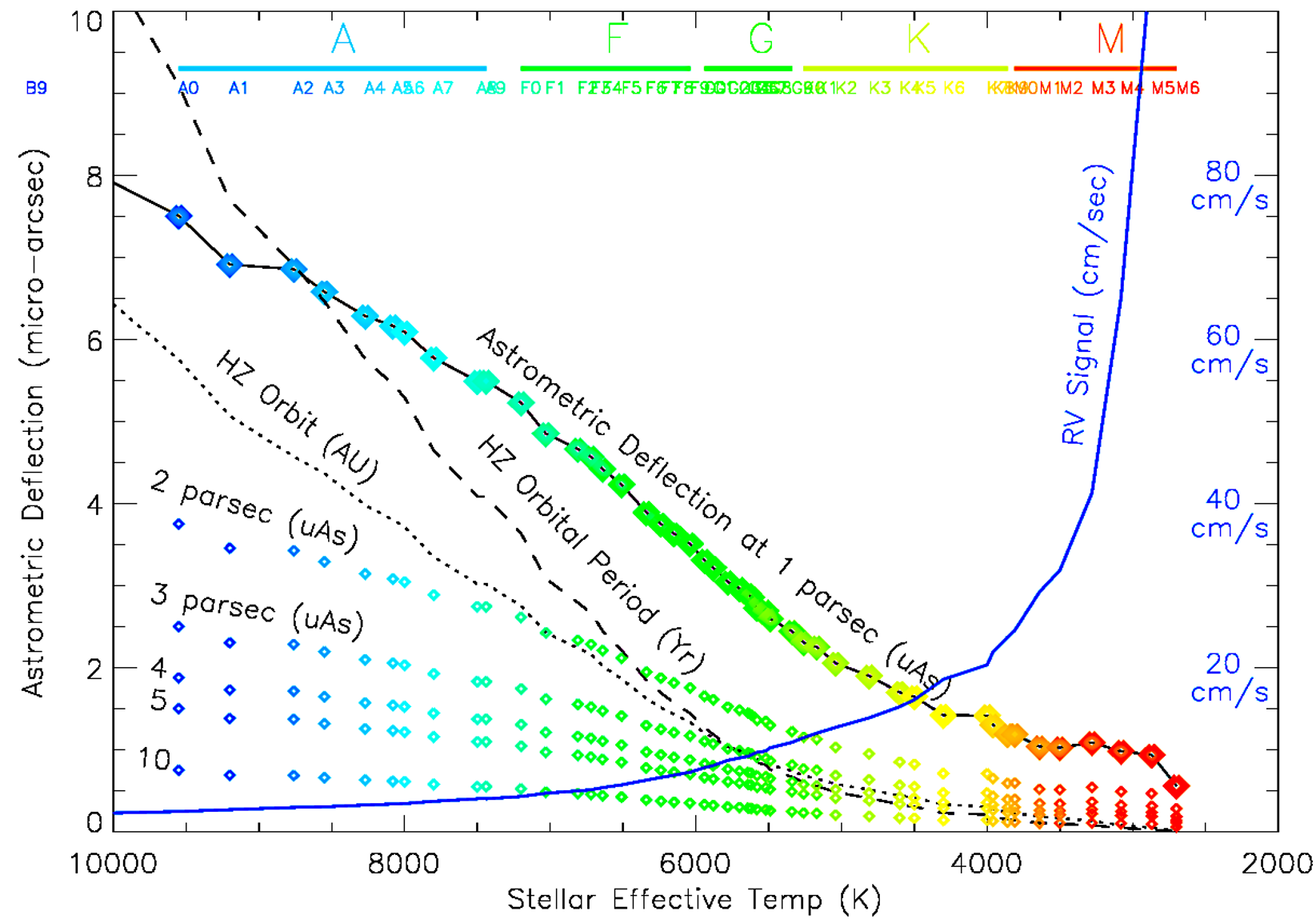}
\caption{The strength of indirect detection signals from an Earth-like exoplanet as a function of stellar effective temperature. Bands encompassing main sequence spectral classes A through M color coded by temperature are indicated at the top. Here, Earth-like is defined as a 1-Earth mass object in an orbit receiving irradiance equivalent to 1\,AU from our Sun (the solar constant). Radial Velocity signals (solid blue line) are seen to exhibit a rapid rise in signal strength (given on the right-hand abscissa) for cooler, M~dwarf systems where they become detectable with present sub-meter per second RV precision. On the other hand, astrometric deflection signals (colored diamond symbols) rise with host star temperature, and decline linearly with distance from values of up to several microarcseconds for the nearest stars. For convenience the plot also indicates the size of the temperate orbit modeled (in AU, left abscissa) and its corresponding orbital period (in years, also left abscissa).}
\label{fig_signalstrength}
\end{figure*}

In contrast to the case for RV, astrometric signal strength in Equation~\ref{eq-astrom} grows with the semi-major axis of the planet orbit, so that it is better suited to probe more distant habitable zones.
While RV has demonstrated success in exploring temperate orbits around M~dwarfs, astrometry is therefore a promising technology for more luminous, Sun-like stars.
Astrometric signals also gain in strength with proximity to Earth, making the technique ideal for exoplanet detection in the near solar neighborhood.
Figure~\ref{fig_signalstrength} visualizes these systematic trends in signal strength with stellar type for both RV (Equation~\ref{eq-rv}) and astrometric detection (Equation~\ref{eq-astrom}).

The most compelling advantage in favor of astrometric detection lies in its (relative) resilience against unavoidable sources of noise arising in the stellar atmosphere. 
Shifts in the apparent location of the center-of-light integrated over the stellar disk occur due to activity such as sunspot groups, faculae and plages moving and evolving on the surface.
Studies employing the Sun as a proxy to estimate surface activity generally find typical levels of intrinsic centroid jitter, projected to the distance of $\upalpha$~Centauri, to be in the range of 0.26--0.65~microarcseconds\cite{Marakov2009,Marakov2010,Lagrange2011,Sowmya2021,Deagan2026jitter}. Over 90-day windows, jitter can fall as low as $0.04\mu\text{as}$ during periods of low activity, and can rise to $0.97\mu\text{as}$ during high activity\cite{Deagan2026jitter}.
So contrary to the situation for RV detection where intrinsic stellar noise is much larger than the desired exoplanet signal, for Astrometry we find signals are a factor of $\sim$a few times larger than the noise.
Furthermore, absolute astrometry has the advantage of being able to determine the companion mass and orbital parameters with fewer of the ambiguities that beset the RV method such as the degeneracy between the orbital inclination and mass.
Recent work\cite{Deagan2024SPIE,Deagan2026surface} shows that astrometric data can also be used to constrain surface properties such as star spots, differential rotation and inclination, holding out the promise that centroid jitter noise might be compensated for with a sufficiently capable stellar model.

Although astrometric detection is one of the longest established methods in all astrophysics -- Le Verrier exploited deviations in the orbit of Uranus to predict the location of Neptune in 1846\cite{1846AN.....25...85L,1846Neptune} -- at the time of writing the NASA exoplanet archive lists only five confirmed exoplanetary discoveries made by astrometric monitoring. 
However astrometric methods now stand on the brink of a major revolution due to data collected by the Gaia mission\cite{Gaia2016} with many thousands of exoplanets expected to be recovered\cite{Perryman2014,Espinoza2023}.
In common with the handful of already confirmed detections, most of these new objects will be giant planets which yield proportionately stronger signals, with discoveries also favoring low-mass host stars\cite{Sozzetti2014}.
Demonstrated outcomes\cite{Gardner2022} from major flagship missions such as Gaia fall short of reaching signal-to-noise floors in the microarcsecond realm enabling discovery of low mass Earth analog systems.

\subsection{Narrow-Angle Astrometry}

Obtaining data with sufficient stability to register the extremely small angular excursions imprinted by an unseen orbiting rocky exoplanet presents a forbidding instrumental challenge.
Much of the difficulty lies in establishing the stability of the frame of measurement -- stable reference points against which the deflections are registered. 
The usual approach is to recover a large enough field of view so that it will be populated by a sufficient number of (relatively) bright background stars to serve this purpose.
This established methodology precipitates immediate challenges: the imaging camera is required to precisely register angles over large ($\gtrapprox$~arcminute) fields, while there is also a contrast problem in simultaneously imaging the (usually bright, nearby) science targets alongside the (usually faint, distant) reference field stars with the same instrument.

One interesting approach that circumvents these problems, albeit for a restricted subset of stellar targets, is narrow-angle astrometric monitoring. 
By targeting only nearby binary or multiple star systems, the requirement to establish a stable frame registered against distant field stars is obviated. 
Instead, the method primarily aims to recover only the \textit{relative} distances between stars within the same (usually binary) system all within a strictly limited ($\sim$ tens of arcseconds scale) science field. 
Furthermore, narrow-angle astrometry is more robust than wide angle astrometry as many classes of systematic error scale with angle\cite{Tuthill2018SPIE}.

While still a challenging task, the precise monitoring of a closely separated pair of stars, both of comparable (or at least not too disparate) brightness, alleviates some difficulties.
One technology proposed to accomplish this task over the years has been long baseline optical interferometry; more specifically phase referenced interferometry\cite{ShaoColavita1992} in which fringes from two stars are tracked, enabling an extremely precise measurement of their angular separation\cite{Lane2004,Kok2013}. 
The primary reason this architecture is so well suited for astrometry is that the signal recovered grows with the scale of the deflection as projected onto the (long) baseline between the telescopes.
An astrometric monitoring program implemented at the longest optical baselines currently available (the CHARA array\cite{CHARA2005}), the ARMADA survey has recently demonstrated measurement precision of around 10~microarcseconds\cite{Armada2021} -- not yet able to detect Earth analogs but sufficient for Neptune-mass planets.

While conferring several advantages as discussed, narrow-angle astrometry also entails some drawbacks. 
As mentioned earlier, only binary (or multiple) star systems can be targeted, and planets recovered must be in circumstellar orbits around one component, not circumbinary orbits. 
Furthermore in common with radial velocity, binary star astrometry introduces blind spots and ambiguities.
Only that component of the gravitational perturbation that modifies the binary separation vector projected onto the plane of the sky will cause an observable deflection in apparent separation. 
Therefore planets on highly inclined orbits (with respect to the binary) will be invisible. 
As the $\alpha$~Cen binary presents a nearly edge-on orbital inclination to Earth ($79^\circ$), most of the blind regions of parameter space corresponds to polar orbits: a fortuitous outcome as such objects are dynamically disfavored from stability arguments anyway.
However (and in common with RV) the ambiguity induced by projection onto the sky plane results in a $\sin(i)$ uncertainty over the exoplanet mass. 
Still another ambiguity arises because any astrometric signals recovered might be caused by an object in orbit around either component of the binary resulting in a pair of degenerate solutions, one for each star.

From an instrumental perspective, a critical challenge for narrow-angle astrometry arises from the removal of data locating background stars.
With no field of predictable reference points populating the image plane, steps must be taken to restore measurement precision against imperfect stability of the optical system (for example, drifts in the focal length) occurring even for the most well designed telescopes. 
The strategy adopted by the \textsc{Toliman} mission entails the integration of an extremely stable physical component into the optics to act as an ``astrometric ruler'' -- a fiducial against which tiny motions can be registered. 
Here this function is performed by a {\it diffractive pupil} whose design and role within the optical system constitutes a major part of the innovation of the program, as detailed in the following sections.

\begin{figure*}
\centering
\includegraphics[width=0.8\linewidth]{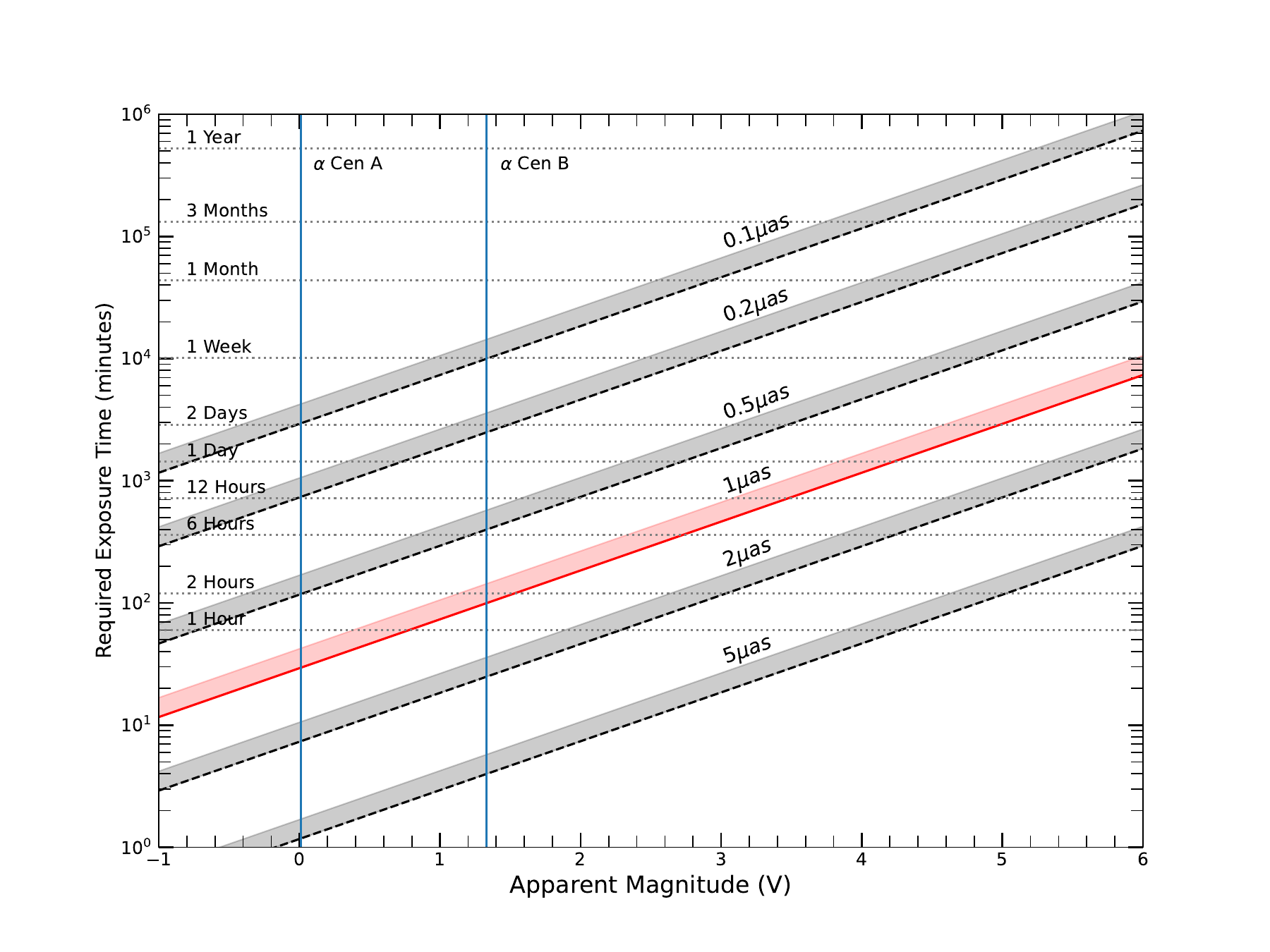}
\caption{The required exposure time for an ideal 125\,mm circular aperture telescope to detect given levels of astrometric signal as a function of magnitude where the only noise processes considered is photon shot noise. The lower boundary of each shaded region shows the required integration time for an ideal Airy disk PSF, while the upper boundary shows the time for the TOLIMAN PSF. The vertical lines correspond to the apparent V-band magnitudes of $\upalpha$~Centauri~A~\&~B, while red highlighting 1\,$\mu$as corresponds to the smallest astrometric signal for any habitable zone Earth-mass object in this system. }
\label{fig_integration}
\end{figure*}

However, prior to discussion of these systems which combat systematic (optical) errors, we must first establish the viability of the measurement against random noise processes fundamental to photon detection.
As quantified in Guyon et al. 2012\cite{Guyon2012}, the positional registration of any image is contingent on the number of photons collected.
Indeed the cost-prohibitive aperture size (meter class) required to observe field stars, driven by exactly this photon noise limit, pushed this mission to a narrow-angle architecture as described above. 
Figure~\ref{fig_integration} depicts the level of astrometric precision that can be accomplished under idealized conditions with a modest-aperture (in this case 125\,mm) telescope.

\subsection{\textsc{Toliman}: science motivation}

Although attaining the levels of astrometric precision required for exoplanetary detection is regarded as a forbidding technical challenge, the manifold threads of technical advantage described so far can all be woven together, and in particular all point favorably for the study of $\upalpha$~Centauri~AB: a very nearby, bright binary comprised of two Sun-like stars.
This confluence of ideas underpins the \textsc{Toliman} mission concept, first proposed in Tuthill et al. 2018\cite{Tuthill2018SPIE}.

As is readily apparent from inspection of Figure~\ref{fig_integration}, it is {\it in principle} possible to get better than 1~microarcsecond astrometric precision from a 125\,mm CubeSat-class telescope in less than a few hours observing provided the target stars are sufficiently bright.
This condition is met by both luminous A and B components of $\upalpha$~Centauri as indicated on Figure~\ref{fig_integration}.
Furthermore as the closest star system to Earth, the astrometric signals are maximized -- deflections from gravitational reflex are calculated to be 2.3\,$\upmu$as and 1.7\,$\upmu$as respectively for an Earth-mass companion in an orbit receiving equivalent insolation to Earth.
Because both stars are bright (V magnitude 0 and 1.3) main-sequence Sun-like stars (Type G2V and K1V), then the $\upalpha$~Centauri AB system boasts two habitable zones featuring orbits of roughly 1 Earth year. 
The apparent separation of the binary also falls within the ideal range for narrow-angle astrometric detection, with its 80\,yr orbit now gradually separating following a close approach (although not periastron) of about 4\,arcseconds in 2015. 

Although M-dwarf Proxima, the third ``C'' component of the $\upalpha$~Centauri triple star system, features at least two known exoplanets\cite{Proxima_2025}, these are neither Earth analogs nor well suited to our methods and are not discussed further. 
Signals recovered using the VISIR instrument on ESO’s Very Large Telescope (VLT) hinted that the system's ``A'' component may host a gas giant class planet orbiting near the habitable zone \cite{AlphaCen_2021}.
Lending considerable support to the original tentative finding, recently analyzed imagery from JWST constrained the planet candidate's properties to a 90--150\,$M_\oplus$ mass object on a $\sim$2--3\,yr orbit \cite{Beichman_2025,Sanghi2025}.
If present, such a massive object will yield astrometric signals two orders of magnitude larger than the rocky planet targets discussed above, yielding both opportunities and challenges.
High precision astrometry can of course fill a valuable role in confirmation of such candidates, and furthermore will help to pin down critical properties of mass and orbit -- even potentially finding exomoons if executed with sufficient precision \cite{Wagner2025}.
However in addition to providing a strong benchmark signal for astrometric missions to confirm, the presence of a high mass planet orbiting on timescales only marginally longer than those corresponding to the habitable zone creates challenges when prospecting for additional lower-mass objects over a monitoring campaign of limited duration. 
Unless the orbital ephemeris of the massive object is very well constrained, signals from any additional lower mass components, particularly those orbiting on similar timescales, will be difficult to disambiguate from extra terms that refine the gas giant orbital parameters.
On the other hand temperate zone planets around $\upalpha$~Cen~B, for example, may have periods as short as 0.4\,yr -- more than a factor of 5 shorter -- separating their signals in frequency space. 

\section{Technologies for precision astrometric measurement}

\subsection{Image registration for astrometry}

There is a substantial literature building information-theoretic principles for image registration; for the purposes of astrometric stellar measurement, Guyon et al.\cite{Guyon2012} showed that the Airy point spread function (PSF) arising from a simple unmodified circular aperture maximizes the image gradient energy and is therefore optimal for image registration.
However the \textsc{Toliman} project has pursued a vigorous program to engineer complex PSFs that spread the starlight spanning the science field. 
There are a number of reasons (discussed below) for this departure from the idealistic Airy case, however the most important is that more complex distributed PSFs can calibrate systematic errors arising from distortions, instabilities, and imperfections in the optical system.
In essence, spreading the light of the two stars into overlapping clouds of sharp speckles enables {\it local} measurements on the sensor in which closely adjacent peaks, each containing light from a different binary component, can be compared.
This renders optical errors common to both, and spans the gap between the two binary components with a calibrated optical ruler. 

Note that this is unlike the case for an Airy function where energy dies off very rapidly away from the bright core.
For such an orthodox optical system (i.e. without a diffractive pupil), relative positional measurements between any two on-sky sources are made with respect to the pixels on the detector, so that recovered signals are susceptible to temporal drifts in the telescope pixel scale. 

The diffractive pupil architecture instead makes measurements with respect to interference fringes cast from structures engraved in the pupil. 
Ideally, this is the first optical element encountered by the starlight, so that instability in optics further downstream will warp the PSFs of all objects in the science field identically (dependent only on the field angle). 
The net result is that the science signal can be isolated from varying optical aberrations delivering measurements robust against mechanical drifts. 

Numerical studies modeling injection/recovery of binary star separations performed over a wide range of noise conditions have vindicated these principles.
In simulation\cite{Desdoigts2025_DiffOptics}, the PSFs arising from complex featured pupils have been shown to dramatically out-perform simple Airy functions whenever sources of systematic noise (such as optical errors like varying defocus) are folded into the scenario.

\subsection{Diffractive pupils for astrometric measurement}

This use of diffraction patterns generated by mechanically stable features engraved in the pupil plane for image registration -- first proposed by Guyon et al.\cite{Guyon2012} -- originally called for a regular array of small opaque dots embedded in the primary mirror. 
When observing a bright star over a narrow optical bandwidth, this results in a diffraction pattern spanning the image plane with a regular grid of sidelobe spots. 
However when considering broadband illumination, bandwidth smearing draws each sidelobe into a spectrum occupying a narrow radial streak or ray. 
Astrometric signal recovery proceeds by registering the location of these rays against background field stars. 
Because the diffractive ruler takes the form of long narrow radial rays, then the positional information recovered must be in the orthogonal ordinate (across the ray, not along it). 
The primary observable then consists of azimuthal positions of (a field of) background stars registered against the nearest diffraction rays\cite{Guyon2012}.

For our case of narrow-angle binary star astrometry, this diffractive pupil formulation has two flaws: (1) it relies on background field stars and (2) it creates a radially-oriented astrometric ruler whose fiducial lines measure {\it azimuthal} shifts but is unable to yield a measurement of the {\it radial} separation of any binary star: TOLIMAN's prime (and only) final observable. 
Where the entire science field consists of only a single pair of stars, formulating a ruler with fiducial lines whose measurement can span the separation between them is essential to obtain the science signal. 
The Guyon et al. formulation of a diffractive pupil is therefore unworkable for our use case motivating consideration of alternative architectures.

\subsection{Reformulating diffractive pupils for precision binary astrometry}

Printed upon materials selected for mechanical stability and extremely low thermal expansion, \textsc{Toliman}'s astrometric ruler is implemented as a diffraction grating located on a transmissive plate at the entrance aperture to the telescope.
Our architecture, while borrowing somewhat in underlying concept from the Guyon\cite{Guyon2012} scheme, is reformulated to the extent that it has little in common with earlier designs.
Starlight is diffracted by a bespoke pupil-plane phase pattern causing it to spread over a narrow portion of the science field around the immediate vicinity of the binary.
Specifically, the resulting PSF pattern is engineered to generate sharp structure oriented to measure {\it radially} separated objects.
Although early conceptual designs explored amplitude apodization, phase gratings were found to be more efficient at spreading light while incurring no throughput penalty. 

The primary design objective for the \textsc{Toliman} diffractive pupil is to spread starlight over a scale corresponding to the science field requiring stabilisation (e.g. less than 10\,arcseconds or so for $\upalpha$~Centauri). 
A further requirement is imposed by the fact that the primary measurement is a {\it radial} distance -- the scalar separation between components of a binary star.
Therefore our task is to engineer the ideal diffraction pattern over which to spread the starlight to optimize our ability to register the locations of the two stars on the image plane sensor (or more specifically, the scalar displacement between them).

\subsection{Design and optimisation of the diffractive phase plate}

The exact form over which starlight is spread in the image plane to form the diffractive ruler will have a profound influence on our ability to recover precise astrometric information.
In the most ideal case, light would be diffracted particularly into the gap separating the two components of the binary star, and the pattern formed would maximize sharp (high gradient energy) structure particularly oriented to form fiducial markers along the line separating the pair of stars.
For operation aboard a space telescope, practical considerations drive the extra requirement that the instrument must observe the binary over the full range of possible orientations (roll angle on the sensor).
This eliminates a large class of possibly interesting patterns (e.g. in which fringe power is asymmetric and directionally oriented with respect to the binary). 
Instead, our primary criterion for optimising the pattern was to emphasize radial gradient energy -- sharp structure encountered as displacement increases from the PSF core.

Engineering phase screens in the pupil plane, particularly those where delays imposed for different regions are in anti-phase, is an effective way to scatter light out of a single bright PSF core into a halo of sharp speckles.
There are a number of ways to implement such phase steps in the pupil, the simplest in concept being to simply sculpt changes in the surface figure of the primary mirror (a strategy that was considered, but expensive to fabricate on a curved polished surface). 
The \textsc{Toliman} project will instead implement the diffraction grating on a custom transmissive flat entrance plate fabricated from a high stability material.
We are exploring several potential pathways to fabricate the desired phase patterning, one of which is a remarkable 2-D printed technology in which a thin coating of liquid crystal is aligned with a scanning laser creating arbitrary holographic phase screens.
Because the technology operates by the manipulation of polarisation, the phase shifts imprinted on the wavefront are intrinsically broadband\cite{Doelman2017,Doelman2020} and the final optic exhibits nearly achromatic performance.

Our task then is to optimize the form of a structured pattern of phase delays imposed at the pupil plane in order to maximize merit functions derived from analysis of the properties of the resultant PSF.
To accomplish this, our group developed the advanced optical simulation code $\partial Lux$ written in the high-performance computer code \textsc{Jax}, which also features the ability to perform automatic differentiation and so deliver gradients computed through even complex sequences of operations.
As has been noted in several domains of science, this property provides a powerful enhancement to our ability to perform numerical optimisation, even where the merit function has a complicated form and the forwards model requires significant effort to compute.

Details of the mathematical and information-theoretic principles that were employed to build codes to optimize the \textsc{Toliman} pupil are described in several publications by Desdoigts et al.\cite{LouisThesis,Desdoigts2024arxiv,Desdoigts2025_DiffOptics}.
It is important to highlight one specific design constraint (for reasons discussed later in Section~\ref{sect:sidelobes}) required the primary phase pattern in the diffractive pupil to be binary valued: that is employing only regions of phase delay of $0$ and $\pi$ throughout.

Another important feature of the \textsc{Toliman} pupil is that it is not inversion-symmetric through the origin; this asymmetry is a requirement for the sign of the even modes of aberration to be uniquely determined in phase retrieval \cite{Martinache_2013,Vortex2024}. This condition (a well known aspect of phase‑retrieval degeneracy\cite{Phasebook}) being satisfied, it is straightforward to do this phase retrieval by gradient descent using our fast \textsc{Jax} modeling code $\partial$Lux \cite{2014JOSAA..31.1348J,2021JOSAB..38.2465W,Desdoigts2023JATIS,Desdoigts2025_DiffOptics} to solve for the entire optical system state including all pupil phase, detector and scene parameters simultaneously.

Critically, these capabilities allow us to optimize the \textsc{Toliman} mask to maximize the information-theoretic content of the images in constraining astrometry, subject to residual uncalibrated phase delays. This is encoded by the Fisher information matrix \cite{fisher1925,Coe2009}, which can be obtained from the matrix of second partial derivatives (the Hessian) of the log-likelihood with respect to the model parameters of interest. In Desdoigts et al. 2024 \cite{Desdoigts2025_DiffOptics}, we directly optimize the pupil phase pattern for \textsc{Toliman} to maximize the constraint on astrometry achievable subject to optical aberrations, and have adopted this design in the manufactured pupil.

\begin{figure}[hbt!]
\centering
\includegraphics[width=1.\linewidth]{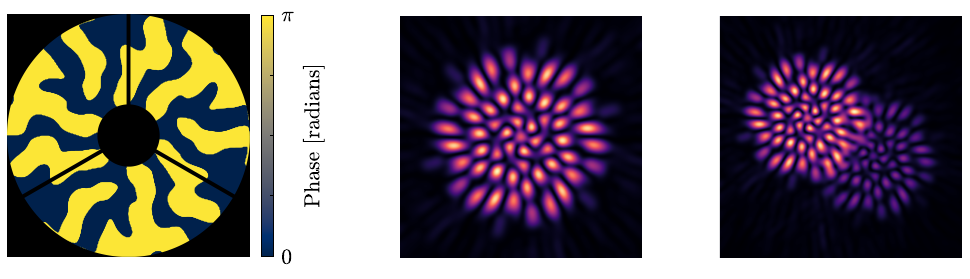}
\caption{\textit{Left:} A depiction of the circular \textsc{Toliman} telescope entrance pupil (including secondary and spider obstructions) together with the final optimized diffractive pupil binary-valued phase pattern. \textit{Middle:} The simulated central region of the resulting PSF. \textit{Right:} A simulation of the science field of $\upalpha$~Centauri binary at the appropriate separation and contrast ratio for A \& B components.}
\label{fig:pupilpsf}
\end{figure}

\subsection{Benefits of coded aperture imaging}

Quite apart from its ability to mitigate errors introduced by instability of the optical surfaces or metering structures, the diffractive pupil offers several other key advantages as noted below.

\begin{itemize}
    \item {\it Mitigation of pixel-based errors.} 
    In contrast to the (much more compact) Airy disk from an open circular pupil, the diffraction pattern spreads information over many pixels. 
    This helps to combat pixel-based noise processes (for example, flat field errors arising from imperfect knowledge of individual pixel gains) because their impact is reduced in proportion to the square root of the number of pixels over which light is spread (assuming statistical independence of such errors).
    
    \item {\it Metrology of the optical system.} 
    The pattern generated by the diffractive pupil constitutes an image-plane wavefront sensor. 
    With sufficient data, errors in tilt and location of all optical surfaces can be recovered, and going to higher orders, a modal representation of distortions on these surfaces.
    This can be accomplished in a number of ways by deployment of various phase retrieval methods.
    The deliberately chosen odd-order symmetry of the diffractive pupil element serves the same function as the asymmetric apodizer in the well established {\it asymmetric pupil wavefront sensor}\cite{Martinache_2013}. 
    Unambiguous recovery of both odd and even order wavefront modes has been demonstrated from diffractive pupil data in simulation and laboratory settings.
    
    \item {\it Characterisation of the sensor.} 
    Calibration of sensors to levels required for precision astrometry requires rigorous and extensive ground-based pre-flight testing.
    However even when perfectly calibrated, environmental damage (for example arising from radiation) will inevitably cause our knowledge of the sensor characteristics to degrade in flight, motivating schemes to implement in-flight monitoring of pixel gains and other characteristics.
    The stable, well-understood nature of the diffraction pattern allows such a calibration scheme to be implemented. 
    Particularly when datasets collected deliver diversity (for example with many pointing dithers), then properties of the pixels such as gains and centroids can be recovered from fitting with large, many-parameter models. 

    \item {\it Reduced effects of nonlinearity and pixel saturation.} 
    Observing very bright stars such as $\upalpha$~Centauri, even telescopes of modest aperture can approach regimes of nonlinear or saturated response for the fastest available readout times of the sensor. 
    Therefore many instruments at modern observatories, both ground and space, are unable to observe bright stars (or must do so through wasteful neutral density filters).  
    In spreading the same flux over several thousand pixels, the diffractive pupil reduces the peak count rate by roughly this proportion, eliminating issues of saturation and constraining pixels to operation in the highly linear low-flux regime.
    
\end{itemize}

\subsection{Stabilisation of the astrometric metrology scale}
\label{sect:sidelobes}

When using the \textsc{Toliman} diffractive pupil for stellar astrometric monitoring, an issue arises because the global scale of the image-plane diffractive ruler is set by three parameters: (1) the physical dimensions of the structure diffracting the light, (2) the focal length (or plate scale) of the imaging system and (3) the effective wavelength of the light.
While the mission is able to control the first of these to acceptable levels (by way of the thermal and mechanical stability of the diffractive plate), variations in the latter two are impossible to distinguish and result in measurement error.
The effective wavelength generating the fringes is subject to variation as the star changes in effective temperature with surface activity, while the plate scale will vary slightly with imperfect stability of the optical metering structure.
The requirement for microarcsecond astrometric precision demands a way to measure and disentangle these two effects: one from the instrument and the other intrinsic to the star. 

This can be achieved by simultaneously monitoring the precise stellar spectrum over the observing band. 
By monitoring the stellar spectral energy distribution, effective temperature changes that drive the effective wavelength variations can be compensated for in the data reduction.

This monitoring function is accomplished by adding a second set of diffractive features to the \textsc{Toliman} pupil.
The addition of a sinusoidal phase grating spanning the pupil will cause the stellar spectrum to appear on the sensor after the manner of a slitless spectrograph. 
Our design implements two such orthogonal gratings, and with judicious choice of both the orientation and cadence of the grating, the stellar spectra are engineered to appear at the four corners of the sensor. 

\begin{figure}
    \centering
    \includegraphics[width=\textwidth]{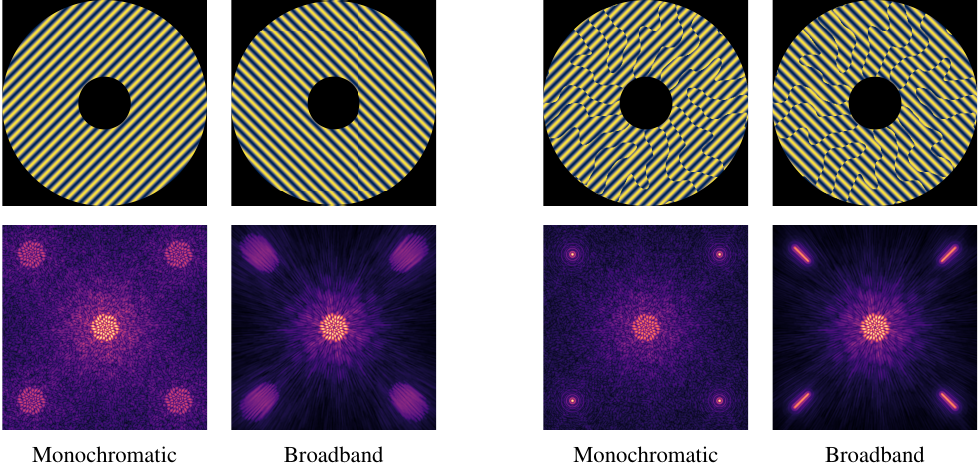}
    \caption{\textit{Left 4 Panels:} simulated pupil and image plane pairs resulting from a simple sinusoidal grating added to the point-design diffractive pupil given in Figure~\ref{fig:pupilpsf}. Upper panels illustrate two sinusoidal phase gratings oriented orthogonally, while the Lower panels show the resulting image plane core/sidelobe features both monochromatically (left) and over a 10\% fractional bandwidth (right). \textit{Right 4 Panels:} same as for the left 4, except that the gratings now have phase inversions at boundary crossings (as discussed in the text) implemented. The color scale is not linear and has been stretched to accentuate faint structure in the images.}
    \label{fig:gratings}
\end{figure}

However, additional design sophistication is required for this system to perform optically as illustrated in Figure~\ref{fig:gratings}.
The straightforward addition of phase gratings to the \textsc{Toliman} diffractive pupil (Figure~\ref{fig:pupilpsf}) does result in the desired 4 sidelobes, however these inherit the form of structured PSF in the core, now echoed at locations in the corners of the sensor.
The problem illustrated in Figure~\ref{fig:gratings} is now readily apparent: the PSF creates a very blurred spectrograph ``slit function'' generating an extremely poor, low-resolution rendering of the stellar spectra.
This makes for an impossible data reduction task to extract the required spectral information in the radial direction.

 The optical design of the diffractive pupil therefore confronts a dilemma. 
Strong diffractive spread of the starlight at the PSF core is required to direct energy into the ruler pattern, however the spectrograph requires that the PSF sidelobes appearing in the corners are as compact (ideally diffraction-limited) as possible.
Fortunately, as illustrated in the four right-hand panels of Figure~\ref{fig:gratings}, there is a way to optimize both.
This is effected by inverting the phase of the sinusoidal grating according to the local phase of the primary diffractive pattern (Figure~\ref{fig:pupilpsf}). 
At boundaries where the primary pattern switches from $0$ to $\pi$ the sinusoidal grating is also switched in phase. 
The effects of doing this are illustrated in Figure~\ref{fig:gratings}, showing it to be remarkably effective at simultaneously accomplishing both of these seemingly incompatible demands.
The sidelobe PSFs are seen to be diffraction-limited Airy functions (optimal for spectrograph), while the central PSF maintains the complex diffractive cloud previously optimised for the binary star astrometric measurement.

Employing this hybrid ``doubly-diffractive'' pupil, we are able to provide contemporaneous spectral monitoring (with $R \sim 700$) of the star and so account for any subtle changes in the effective wavelength.

\subsection{Direct plate scale recovery from spectral sidelobes}

Having now incorporated a spectrograph into the optical design, stellar spectral lines will therefore be imprinted into the sidelobes that will also give an additional, and more direct, way to monitor the plate scale.
Arising from basic atomic physics, these lines provide an absolute wavelength reference which, in combination with the (presumed) mechanical stability of the diffractive pupil, locates these lines in the image plane allowing us to fully specify the plate scale of the entire instrument. 
In an optical arrangement that places the stellar spectra at the corners of the detector a few arcminutes from the PSF core, a geometrical amplification of any changes in the focal length is created greatly augmenting the sensitivity to plate scale.

\begin{figure}[hbt!]
\centering
\includegraphics[width=1\linewidth]{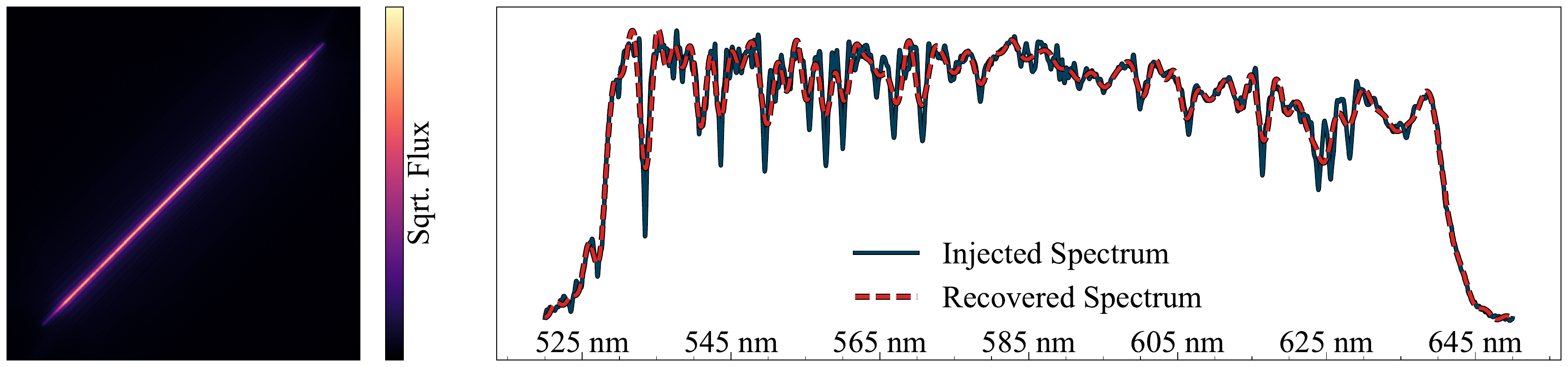} 
\caption{\textit{Left:} One of the four sidelobes of the \textsc{Toliman} pupil, displaying intensity variations that encode the stellar spectrum. \textit{Right:} The injected and recovered spectra of this sidelobe, illustrating the resolving power of the \textsc{Toliman} spectrograph.}
\label{sidelobe}
\end{figure}

Therefore the doubly-diffractive pupil with its integrated spectrograph provides two entirely independent calibrations: both the stellar spectral energy distribution (and therefore effective wavelength) as well as the image scale in absolute units referenced against stellar spectral lines. 
Information provided by these two pathways is not independent providing a way to cross-check the calibration at sub-microarcsecond scales.

\section{The \textsc{Toliman} Space Telescope Hardware}

\subsection{The \textsc{Toliman} mission: overview}

\renewcommand{\thefootnote}{$\dagger$}

The \textsc{Toliman}\footnote{ Toliman, the International Astronomical Union (IAU) name for $\upalpha$~Centauri~B, stands for Telescope for Orbit Locus Interferometric Monitoring of our Astronomical Neighborhood.}
mission\cite{Bendek2018} will fly a 125\,mm clear aperture optical telescope hosted aboard a 16U CubeSat bus in low Earth orbit with a primary goal to monitor the $\upalpha$~Centauri~AB system for a period of three years. 
The optical system has been deliberately kept as simple as possible with only two powered surfaces, with high stability materials (Zerodur and silicon carbide) chosen for most optical components and the telescope metering structure.
The entrance face to the telescope hosts the diffractive pupil that performs the role of the astrometric ruler as discussed in earlier sections.

Observation from low Earth orbit sidesteps ground-based angular errors arising from turbulent atmospheric seeing, while also providing a relatively benign operating environment to stabilise the telescope (with no gravity loading for example).
Levels of spaceflight thermal stability can be superior to those on Earth\cite{REFFERT2009329}, although accomplishing high levels of stability is more challenging in (less expensive) lower orbits where varying radiative loads are unavoidable. 

\subsection{\textsc{Toliman} subsystems and components}

This section presents a brief sketch of the main components and parameters for the \textsc{Toliman} mission, although detailed description lies beyond the scope of the present paper.

The telescope's primary mirror and metering structure are all manufactured from silicon carbide by Aperture Optical Sciences in the USA.
Although the simplest possible optical design would have the diffractive pupil implemented on the primary mirror, fabricating such a patterned structure on a curved optic proved costly and technically challenging. 
Instead it will be implemented on a flat transmissive plate mounted at the front entrance to the telescope made of a very high stability material such as Zerodur or ULE Glass.
Several fabrication pathways for printing or engraving the required phase pattern onto this bespoke optic are being explored, with the most promising to date entailing a laser-written geometrical phase grating patterned onto a liquid crystal coating\cite{HAM_2021}. 
This is subsequently sandwiched with a second plate in a glued assembly whose other surfaces carry interference filter coatings to restrict observations to a bandwidth spanning approximately 530 to 640 nanometers.
This bandpass was chosen for its relative cleanliness from strong stellar lines, in particular upper atmosphere or chromosphere features which might cause variation in signals.
Details of the design and manufacture of this optical element can be found in Langford et al. 2024\cite{Langford2024SPIE}.
After the diffractive pupil plate, the light encounters only two powered surfaces -- the primary and secondary mirrors -- comprising a Ritchey-Chr\'{e}tien optical system that forms images with an $f$-ratio of 10.5 onto a sensor from the Sony IMX family (exact model has not yet been finalized) expected to have $\sim 2.5\upmu$m pixels leading to a sampling of about 3 pixels per fringe for the highest expected spatial frequencies.

Several operational requirements challenge the normal levels of performance obtained by CubeSat-class missions.
Pointing accuracy and stability is among the foremost of these. 
Science requirements drive the relatively stringent targets of absolute pointing to within a couple of arcseconds and a jitter/drift rate of better than about 1 arcsecond per second.
Such requirements flow down from studies of the impact of motion blur on the final science productivity, as we have quantified elsewhere\cite{Charles2025}.
As commercial CubeSat star tracking systems available to us were not able to meet this specification, the telescope is integrated with an active system able to point independently of the spacecraft bus over a few tens of arcseconds. 
This tip-tilt system employs piezo-electric actuators that enable a limited range of fine motion control of the science payload. 

A second challenge lies in ensuring the thermal stability of the telescope against the varying radiative loads arising in low Earth orbit. 
This is accomplished by way of several strategies including baffles and passive insulation, maneuvering in orientation with orbit, active heaters and heat straps connected to external radiators. 
Thermal studies of the spacecraft detailing these will be discussed in a forthcoming publication.

\subsection{\textsc{Toliman} data handling and reduction}

This section is also restricted to a brief overview of material discussed and presented elsewhere.
Sensor frames will be recorded at a rate of about 10\,Hz and, without compression, would generate a data stream that would be too costly to downlink. 
Fortunately information in each frame is sparsely distributed over a small subset of pixels, notably the science field containing the binary near the center of the chip, and the four sidelobes in the corners.
Our first strategy in data compression is simply to excise these relevant regions and discard the majority of sensor pixels containing no light.
After further lossless compression, data will be downlinked for subsequent analysis.

The task of recovering the extremely small signals presents daunting challenges; any algorithm or code to accomplish this must have its foundations in statistically principled treatments of the data.
As for the optical design tasks described earlier, our approach\cite{Desdoigts2022SPIE} is founded upon physically informed forward modeling of the signal chain from the star through to the sensor enabled by the fast, efficient and differentiable language \textsc{Jax}. 
The speed and accuracy of this simulation code, combined (as before) with the availability of differential computations, enables forward model fitting\cite{Desdoigts2023JATIS} as well as machine learning methods\cite{Veneri_2021} to be deployed.
In simulation, injection/recovery of signals at the required levels of precision has been demonstrated, while work to inject ever more realistic sources of noise and systematic error into these tests is now underway.
Outcomes will be detailed in forthcoming published work. Importantly, the same code used to \textit{design} the Toliman pupil \cite{Desdoigts2025_DiffOptics} can be fine-tuned on in-flight data to be a digital twin of the actually-existing instrument as built, and used as the forward model in Bayesian inference of the science data\cite{LouisAmigo25,MaxAmigo25}. 

\subsection{Error budget}

Furnishing a fully fleshed out error budget entails analysis of the flowdown of expected levels of all aspects of instrument performance into their impact on the final mission observables. 
Such a major undertaking is beyond the scope of this paper (indeed, several sub-components furnish material for forthcoming papers), and furthermore several aspects of performance are not yet final creating uncertainty in the inputs.
However, to give a rough sketch of the anatomy of an error budget for a TOLIMAN-like instrument, we present the early design-stage (aspirational) breakdown of terms targeting delivery of a final precision better than 1\,$\mu$as. 

The error budget for a precision narrow-angle astrometric mission can be structured into a hierarchy of six primary categories: photon noise, plate scale calibration, detector errors, pointing and jitter, astrophysical noise, and geometric/relativistic effects. 
Photon noise represents the fundamental statistical uncertainty in photon centroiding and is allocated 0.15\,$\mu$as, driving the exposure time requirement to approximately 2 hours for Alpha Cen AB. 
Plate scale calibration accounts for angular measurement scale precision, where an allocation of 0.3\,$\mu$as drives thermal stability requirements of the optical system ($\sim$ ±0.1°C), diffractive pupil variations, and the performance of the spectral sidelobe calibration subsystem. 
Detector errors, encompassing intra-pixel and inter-pixel Quantum Efficiency (QE) variations, physical pixel grid offsets and various sources of non-ideal sensor performance are allocated 0.25\,$\mu$as assuming pixel photocenter knowledge exceeds 0.001 px; these errors bias the Point Spread Function (PSF) photocenter relative to pixel registration, coupling this category with pointing stability. 
Pointing and jitter account for spacecraft instabilities that cause PSF smearing and wavefront errors (WFE) due to beam walk, requiring a pointing accuracy at the $\sim$ arcsecond level and jitter below 1”/s to achieve an estimated error of 0.3\,$\mu$as. 
Astrophysical noise, including the influence star spots (see Section~\ref{sec:astro_mon}) and background stars, is allocated 0.6\,$\mu$as. 
Finally, geometric and relativistic effects are allocated 0.15$\mu$as based on gnosis of spacecraft velocity (at the 1 m/s level) together with milliarcsecond-level precision of the binary orbit and angle currently constrained by GRAVITY and ALMA. 
The resulting aggregate single-visit error is 0.8 $\mu$as, which scales inversely with the square root of the number of observations for uncorrelated noise sources.
In practice, keeping error terms within all these limits, particularly where sources of systematic noise from various sources of instability are present, will present challenges.

\subsection{Status and next steps}

At the time of writing, all major mission components are either under fabrication or at an advanced stage in design.
The SiC telescope and metering structure is nearly complete, while the 16U spacecraft bus together with all required flight subsystems is fully designed with fabrication and assembly underway. 
We plan to complete integration and all testing in readiness for a 2027 launch window on a Falcon~9 rideshare, although we are unable to forecast the scheduling of insertion into our (strongly desired) dusk-to-dawn orbit (chosen for superior thermal stability).
The data processing pipeline architecture and operations is currently being developed at the SETI Institute.
Major mission partners include Breakthrough Watch (Breakthrough Initiatives, sponsored by the Breakthrough Prize Foundation), the University of Sydney, Australian Research Council, and the SETI Institute. Total funding amounts to approximately 5M USD. 
Furnishing all mission components including bus, launch and mission ops for a high performing 16U spacecraft within this envelope has driven innovation and new thinking from the TOLIMAN team.

\section{Conclusions}

This paper has discussed a suite of technological innovations that underpin the \textsc{Toliman} mission.
This constitutes a framework of optical hardware and signal processing to empower the delivery of microarcsecond astrometric precision, as required for habitable-zone Earth-analog exoplanet detection. 
The project aims to accomplish this within the envelope of a modest-aperture space telescope mission flying aboard a CubeSat-class bus. 
Even where novel strategies to mitigate error and noise are in place, the working parameters of the mission still place stringent demands upon many subsystems (e.g. thermal control and telescope pointing stability), many of which are particularly challenging from a CubeSat platform in low Earth orbit.  
As data over several orbital cycles will be required, the mission plans for a minimum duration of 3 years to deliver the final science yield. 
The project is now entering final fabrication stages\cite{Tuthill2024SPIE} in anticipation of 2026 integration of the payload and bus for launch readiness.
Our program has also had the intent to raise the profile and technological readiness levels of the science and instrumentation for relative astrometry at extreme precision, with implications for several missions and projects presently at planning stages.
Among the programs that would benefit from TOLIMAN's advancement of diffractive pupil technologies, the SHERA mission from Caltech/JPL\cite{Shera2025AAS} has perhaps the closest match in scope, although potentially outcomes are relevant to a number of future astrometric missions such as the proposed European Theia telescope\cite{Theia2022} under consideration for European Space Agency support.  


 \section{Data and Code availability}
 
 Codes employed for simulations and modeling of telescope performance are available on the $\partial$Lux GitHub: \href{https://github.com/LouisDesdoigts/dLux}{github.com/LouisDesdoigts/dLux}.

\acknowledgments 

The original scientific impetus as well as critical financial and logistical support for this research program were provided by Breakthrough Watch, managed by the Breakthrough Initiatives and sponsored by the Breakthrough Prize Foundation. The work builds upon an earlier feasibility study jointly by the Breakthrough Initiatives and the University of Sydney, with further resources from the Australian Research Council grant LP210200594. 
The authors acknowledge the technical assistance provided by the Sydney Informatics Hub, a Core Research Facility of the University of Sydney.
The authors acknowledge the support by the SETI Institute for mission operations.

The authors acknowledge the Gadigal people of the Eora Nation upon whose ancestral lands The University of Sydney now stands. We pay our respects to Elders past and present, and extend that respect to all Aboriginal and Torres Strait Islander Australians.
BJSP was funded by the Australian Government through the Australian Research Council DECRA fellowship DE210101639. He acknowledges the traditional custodians of the Macquarie University land, the Wallumattagal clan of the Dharug nation, whose cultures and customs have nurtured and continue to nurture this land since the Dreamtime, and pays his respects to Elders past and present.

\section{Disclosure Statement}

The authors declare there are no financial interests, commercial affiliations, or other potential conflicts of interest that have influenced the objectivity of this research or the writing of this paper.


\bibliography{references}   
\bibliographystyle{spiebib} 

\end{document}